\documentclass[preprint,12pt]{elsarticle}

\usepackage{amssymb}
\usepackage{bm,color,bbm}

\usepackage{ulem} 

\usepackage{hyperref, mathtools,graphicx,amsmath}

\usepackage{cases}
\usepackage{natbib}

\journal{Physics Letters A}

\newcommand{\beq}{\begin{equation}}

\newcommand{\eeq}{\end{equation}}

\newcommand{\bqa}{\begin{eqnarray}}

\newcommand{\eqa}{\end{eqnarray}}

\newcommand{\nn}{\nonumber}

\newcommand{\rt}[1]{\sqrt{#1}\,}

\newcommand{\bra}[1]{ \langle{#1} |}

\newcommand{\ket}[1]{ |{#1} \rangle}

\newcommand{\sq}[1]{\left[ {#1} \right]}

\newcommand{\an}[1]{\left\langle{#1}\right\rangle}

\newcommand{\tr}[1]{{\rm Tr}\sq{ {#1} }}

\newcommand{\mf}{\mathbf}

\begin{document}

\begin{frontmatter}



\title{Monogamy relations of nonclassical correlations for multi-qubit states}


\author[first]{Shuming Cheng\corref{cor1}}
\ead{shuming.cheng@griffithuni.edu.au}
\author[second]{Lijun Liu\corref{cor2}}
\ead{lljcelia@126.com}

\cortext[cor1]{Corresponding author}
\cortext[cor2]{Corresponding author}

\address[first]{Centre for Quantum Computation and Communication Technology (Australian Research Council), Centre for Quantum Dynamics, Griffith University, \\ Brisbane, QLD 4111, Australia}

\address[second]{College of Mathematics and Computer Science, Shanxi Normal University, \\Linfen 041000, People's Republic of China}

\begin{abstract}
Nonclassical correlations have been found useful in many quantum information processing tasks, and various measures have been proposed to quantify these correlations. In this work, we mainly study one of nonclassical correlations, called measurement-induced nonlocality (MIN). First, we establish a close connection between this nonlocal effect and the Bell nonlocality for two-qubit states. Then, we derive a tight monogamy relation of MIN for any pure three-qubit state and provide an alternative way to obtain similar monogamy relations for other nonclassical correlation measures, including squared negativity, quantum discord, and geometric quantum discord. Finally, we find that the tight monogamy relation of MIN is violated by some mixed three-qubit states, however, a weaker monogamy relation of MIN for mixed states and even multi-qubit states is still obtained.
\end{abstract}


\end{frontmatter}

\section{Introduction}\label{Introduction}

Quantum correlations have been not only recognized as fundamental properties in the quantum regime that depart from the classical world, but also regarded as useful resources in numerous quantum information and quantum computation tasks~\cite{NC00}. Thus, it is a prime task in quantum information theory to characterize and quantify these nonclassical resources~\cite{HHHH09,MBCPV12,BCPSW14}.

One of fundamental features in quantum world is the existence of quantum nonlocality, especially Bell nonlocality~\cite{BCPSW14}, signaling distinct incompatibility between quantum mechanics and local realism~\cite{B64}. In particular, Bell nonlocality could be revealed in the very simple scenario of two-qubit systems, shared by distant observers, where each observer chooses one of two dichotomic measurements on each qubit~\cite{CHSH69}. Moreover, it was found that the Bell nonlocal states, such as the Einstein-Podolsky-Rosen state~\cite{EPR35}, are capable of accomplishing jobs impossible in the classical world, such as device-independent quantum key distribution (QKD)~\cite{E91,AGM06}, quantum teleportation~\cite{Tele93}, and super-dense coding~\cite{BW92}. 

Recently, another nonlocal effect, called measurement-induced nonlocality (MIN), was introduced by Luo and Fu in~\cite{LF11}. This nonlocal effect is more general than Bell nonlocality and describes the global effects caused by the local measurements on one side~\cite{LF11}. In this work, we explore the potential relationships between these two kinds of nonlocality. Furthermore, they are also compared to the well-known quantum entanglement, and to nonclassical correlations beyond entanglement, such as quantum discord~\cite{OZ01,LV01} and geometric quantum discord~\cite{DVB10,LF10}. 

Another peculiar quantum feature is the monogamy of quantum correlations which constrains the distribution of quantum correlations among multiparty systems. For example, when Alice and Bob share a maximally entangled state~$\frac{1}{\rt{2}}(\ket{00}+\ket{11})$, then each party can only be classically correlated with the third party with no entanglement at all. This phenomenon is termed monogamy, and for Bell nonlocality ensures the security of quantum cryptographic protocols~\cite{E91}. Monogamy relations are already known for concurrence~\cite{CKW00,OV06}, negativity~\cite{OF07}, Bell nonlocality~\cite{SG01,TV06,CH17}, quantum steering~\cite{R13,MJJWR14,CMHW16,XKAH17}, and quantum discord~\cite{SAPB12,BZYW13}. 

Along this line, we are interested in whether MIN obeys a similar monogamy relation. We give an affirmative answer for pure three-qubit states, and thus disprove claims in Ref.~\cite{ADA12} that MIN does not satisfy such a monogamy relation. Generally, we show that three-qubit states and arbitrary $n$-qubit states obey another kind of monogamy relations of MIN.

This paper is structured as follows. In Sec.~\ref{Preliminaries}, we  introduce the basic definitions required in the two-qubit scenario and show that MIN is no larger than the Horodecki parameter~\cite{HHH95}, which quantifies the maximal violation of a Bell inequality. Then, we derive a tight monogamy relation of MIN for pure three-qubit states in Sec.~\ref{3-qubit}, and also recover known monogamy relations for negativity, quantum discord, and geometric quantum discord as byproducts. In Sec.~\ref{multiqubit}, a counterexample is constructed to disprove the universality of monogamy of MIN for more general cases, including mixed three-qubit states and multi-qubit states, but an alternative form of monogamy relations for general states is still obtained. Finally, we conclude with discussions in Sec.~\ref{Discussion}.

\section{Measurement-induced nonlocality v.s. Bell nonlocality}\label{Preliminaries}

A two-qubit state $\rho_{AB}$ shared by Alice and Bob can be written as

\begin{align}
\rho_{AB} =  \frac{1}{4} \bigg(  I_A \otimes I_B+{\mf{a}} \cdot \boldsymbol{\sigma} \otimes I_B+I_A \otimes {\mf{b}} \cdot \boldsymbol{\sigma} +\sum^3_{j,k=1}T_{jk}\,\sigma_j\otimes\sigma_k \bigg). 	\label{state}
\end{align}
Here, $\boldsymbol{\sigma}\equiv (\sigma_1, \sigma_2, \sigma_3)$ refers to the vector of Pauli spin operators. $I_A$ and $I_B$ are identity operators. ${\mf{a}}$ and ${\mf{b}}$ correspond to the Bloch vectors of Alice's and Bob's reduced states, and $T$ is the spin correlation matrix with $T_{jk}=\langle\sigma_j\otimes \sigma_k\rangle$. Complementary to the locally accessible information $\mf{a}$ and $\mf{b}$, the spin correlation matrix $T$ is of great importance in encoding the global information and the strength of the quantum correlations of the
qubits~\cite{HHHH09,MBCPV12,BCPSW14}. 

To measure the nonlocal effects induced by local measurements on one side, Luo and Fu proposed the measurement-induced nonlocality~\cite{LF11}
\beq
\mathcal{D}^{A\rightarrow B}_M:= \max_{\Pi^A} ||\rho_{AB}-\Pi^A(\rho_{AB})||^2, \label{measurement}
\eeq
where the maximum is taken over all von Neumann measurements $\{\Pi^A_j\}$ that preserve Alice's local state, i.e., $\Pi^A(\rho_A) :=\sum_j\Pi^A_j\rho_A\Pi^A_j=\rho_A$.  $||X ||:=(\tr{X^\dagger X})^{1/2}$ denotes the Hilbert-Schmidt norm, and the notation $A\rightarrow B$ specifies Alice as the measuring party. Similarly, the nonlocality induced by Bob's local measurements $\mathcal{D}^{B\rightarrow A}_M$ could also be defined. Interestingly, it was proven in~\cite{LF11} that MIN has asymmetric property, i.e., $\mathcal{D}^{A\rightarrow B}_M\neq\mathcal{D}^{B\rightarrow A}_M$.

For an arbitrary two-qubit state $\rho_{AB}$, the MIN admits an explicit form~\cite{LF11}
\begin{numcases}
{\mathcal{D}^{A\rightarrow B}_M = }
{ \frac{1}{4}\left(\tr{T T^\top}-\frac{1}{a^2}\mf{a}^\top T T^\top\mf{a} \right) } & $\mf{a}\neq \mf{0}$ \label{aneq0},  \\
{\frac{1}{4}\left(\tr{T T^\top}-s_3\right) } & $ \mf{a}= \mf{0}$ \label{aeq0}. 
\end{numcases}
Three eigenvalues $s_1, s_2, s_3$ of the symmetric matrix $T T^\top$ are arranged in descending order, i.e., $s_1\geq s_2 \geq s_3\geq 0$, and here and elsewhere we use $x=|\mf{x}|$ to represent the modulus of a vector $\mf{x}$.  Obviously, the MIN of a state lies in the interval $[0, \frac{1}{2}]$ and achieves its maximum value if and only if the state is locally unitary equivalent to any Bell state~\cite{LF11}. Other basic properties of MIN have been listed in~\cite{LF11}. However, this MIN based on the Hilbert-Schmidt (HS) norm suffers from one weakness that it may increase under the completely-positive and trace-preserving (CPTP) maps on Bob's side~\cite{M00,MH15}. To overcome this weak point,  other MINs are proposed, such as based on the trace-norm~\cite{MH15}, the relative entropy~\cite{XWL12}, the fidelity~\cite{MS17}, and the two-sided projective measurements~\cite{G13}. The basic properties of these MINs are referred to Refs.~\cite{MH15,XWL12,MS17,G13}. In this work, we explore the connections between the MIN based on the HS-norm and other quantum correlations and investigate the distribution of MINs for three-qubit states.

Further, Bell nonlocality characterizes whether the outcome statistics generated by local measurements on both sides could be explained by a local hidden variable theory. This nonlocal effect could be exposed in the very simple scenario of two-qubit systems, shared by distant observers, where each observer involves two dichotomic measurements on each qubit. Specifically, each party has two measurements with outcomes $+1$ or $-1$, and these binary measurements are assumed to be Hermitian operators: $A_1=\mf{a}_1\cdot \boldsymbol{\sigma}, A_2=\mf{a}_2\cdot \boldsymbol{\sigma}$ for Alice and $B_1=\mf{b}_1\cdot \boldsymbol{\sigma}, B_2=\mf{b}_2\cdot \boldsymbol{\sigma}$ for Bob. Then, Bell nonlocality is witnessed by violating the Bell-Clauser, Horne, Shimony, and Holt (CHSH) inequality~\cite{CHSH69}
\beq
\an{\mathcal{B}}^2=(\tr{\mathcal{B}\rho})^2\leq 4,
\eeq
with the Bell operator $\mathcal{B}=A_1\otimes B_1+A_1\otimes B_2+A_2\otimes B_1-A_2\otimes B_2$. 

A two-qubit state $\rho_{AB}$ is Bell nonlocal if it violates the Bell-CHSH inequality for some set of local measurements. It is remarkable that the above statement still holds even if the state is not limited to qubit states or even does not admit a quantum description. Hence, determining whether $\rho_{AB}$ is Bell-CHSH nonlocal or not is equivalent to check if~\cite{HHH95}
\beq 
\mathcal{M}:=\frac{1}{4}\max_{A_1, A_2, B_1, B_2} \an{\mathcal{B}}^2= s_1+s_2=\tr{T T^\top}-s_3\leq 1 \label{Horodecki}
\eeq
Here the Horodecki parameter is denoted as $\mathcal{M}$, quantifying the maximal violation of Bell-CHSH inequality.

It follows immediately from Eqs.~(\ref{aeq0}) and~(\ref{Horodecki}) that there is
\beq
\mathcal{D}^{A\rightarrow B}_M=\frac{1}{4}\mathcal{M},~~~ \mf{a}=\mf{0}.
\eeq 
For $\mf{a}\neq \mf{0}$, note that $s_3\leq \mf{n}^\top T T^\top\mf{n} \leq s_1$ for an arbitrary unit vector $\mf{n}$. Hence, choosing $\mf{n}=\mf{a}/a$,  yields from Eq.~(\ref{aneq0}) that 
\beq
\mathcal{D}^{A\rightarrow B}_M=\frac{1}{4}\left(\tr{T T^\top}-\mf{n}^\top T T^\top\mf{n} \right)\leq\frac{1}{4}\mathcal{M}. \label{MINH}
\eeq
for two-qubit states generally. This implies that the Horodecki parameter provides an upper bound for MIN. Since $\mathcal{M}$ is symmetric under the interchange of Alice and Bob,  one similarly has $\mathcal{D}^{B\rightarrow A}_M\leq\frac{1}{4}\mathcal{M}$, corresponding to the choice $\mf{n}=\mf{b}/b$.

Finally, it is of interest to also consider other nonclassical correlations. For example, in contrast to Eq.~(\ref{measurement}), geometric quantum discord is defined as~\cite{DVB10}
\beq
\mathcal{D}^{A\rightarrow B}_G:=\min_{\Pi^A} ||\rho_{AB}-\Pi^A(\rho_{AB})||^2, \label{geometric}
\eeq
where the minimum is taken over all von Neumann measurements. It is apparent that $\mathcal{D}^{A\rightarrow B}_G\leq \mathcal{D}^{A\rightarrow B}_M$.  Additionally, for two-qubit states, it was shown in~\cite{GA11,GA11B} that the geometric quantum discord further upper bounds both the computable entanglement measure $\mathcal{N}$~\cite{VW02} and quantum discord $\mathcal{D}$~\cite{OZ01,LV01}. Thus, we are able to obtain an ordering chain of these nonclassical  correlation measures
\beq
\mathcal{N}^2, (\mathcal{D}^{A\rightarrow B})^2\leq 2\mathcal{D}^{A\rightarrow B}_G\leq 2\mathcal{D}^{A\rightarrow B}_M \leq\frac{1}{2} \mathcal{M}, \label{ordering}
\eeq
for two-qubit states.  This ordering chain is useful in exploring the monogamy phenomenon for multi-party systems. In particular, if there exist monogamy relations for correlation measures in the right side of above ordering~(\ref{ordering}), such as MIN or Bell nonlocality, it is highly possible that these correlation measures in the left side also obey a similar monogamy relation. We will show it is indeed the case in the next section.

\section{ Monogamy relations for pure three-qubit states}\label{3-qubit}

Consider a  general pure three-qubit state of a tripartite system held by Alice, Bob, and Charlie:
\beq
\ket{\phi_{ABC}}=\sum_{i,j,k=0}^1a_{ijk}\ket{ijk}. \label{3qubit}
\eeq 
In this section, we study the problem whether there exists a monogamy relation 
\beq
 \mathcal{E}^{A: B}+\mathcal{E}^{A: C}\leq \mathcal{E}^{A: BC},
\eeq
for pure three-qubit states. Here $\mathcal{E}^{A: B}$ is the nonclassical correlation measure between Alice and Bob, $\mathcal{E}^{A: C}$ is between Alice and Charlie while $ \mathcal{E}^{A: BC}$ is between Alice and the joint Bob and Charlie.

First note that the reduced bipartite states $\rho_{AB}={\rm Tr}_C[\ket{\phi_{ABC}}\bra{\phi_{ABC}}]$ and $\rho_{AC}={\rm Tr}_B[\ket{\phi_{ABC}}\bra{\phi_{ABC}}]$ can be expressed similarly to Eq.~(\ref{state}). Moreover, if a $2\times n$ quantum system is in a pure state, then it can be transformed into a pure two-qubit state under local unitary operations. As a consequence, for the pure three-qubit state $\ket{\phi_{ABC}}$ in Eq.~(\ref{3qubit}), the nonclassical correlations shared by Alice and the rest of the parties could be quantified easily as one has~\cite{LF11,DVB10,HHH95,VW02}
\beq
(\mathcal{N}^{A:BC})^2=2\mathcal{D}^{A\rightarrow BC}_G=2\mathcal{D}^{A\rightarrow BC}_M=\mathcal{M}^{A:BC}-1=1-a^2,\label{identity}
\eeq
while the quantum discord coincides with entropy of entanglement and is given~\cite{GA11}
\beq
\mathcal{D}^{A\rightarrow BC}=h(a):=-\frac{1-a}{2}\log\frac{1-a}{2}-\frac{1+a}{2}\log\frac{1+a}{2}.
\eeq

\subsection{Pure states with $\mf{a}=\mf{0}$} \label{purea0}

For the case where Alice's qubit is in a maximally mixed state, Eq.~(\ref{identity}) immediately yields
\beq
4\mathcal{D}_M^{A\rightarrow BC}=\mathcal{M}^{A:BC}=2. \label{maxi}
\eeq
Since it has been proven in~\cite{CH17} that there is a tight monogamy relation for Bell-CHSH nonlocality, 
\beq
\mathcal{M}^{A:B}+\mathcal{M}^{A:C} \leq 2=\mathcal{M}^{A:BC}, \label{monoM}
\eeq
where the final equality derives from Eq.~(\ref{maxi}), with Eq.~(\ref{MINH}), this immediately leads to 
\beq
\mathcal{D}_M^{A\rightarrow B}+\mathcal{D}_M^{A\rightarrow C} \leq \frac{1}{2}=\mathcal{D}_M^{A\rightarrow BC}. \label{monoMIN}
\eeq
Hence, we obtain analytically a monogamy relation for MIN when $\mf{a}=\mf{0}$. This relation is tight because the state $\frac{1}{2}\left(\ket{010}+\ket{011}+\ket{100}+\ket{101} \right)$  saturates both the monogamy relation~(\ref{monoM}) for Bell nonlocality and the monogamy relation~(\ref{monoMIN}) for MIN. We remark that the monogamy relation~(\ref{monoMIN}) disproves the claim and numerical results in~\cite{ADA12}.

\subsection{General pure three-qubit states}\label{pure}

When $\mf{a}\neq \mf{0}$, the equality $\mathcal{M}^{A:BC}=2$ in Eq.~(\ref{monoM}) does not hold any more and hence the monogamy relation $\mathcal{M}^{A:B}+\mathcal{M}^{A:C} \leq \mathcal{M}^{A:BC}$ could be violated generally. For example, given the state $\alpha\ket{000}+\beta\ket{111}$ with $|\alpha|\neq|\beta|$, we have  
\beq
\mathcal{M}^{AB}+\mathcal{M}^{AC}=2 > \mathcal{M}^{A:BC}=2-(|\alpha|^2-|\beta|^2)^2.
\eeq

However, for MIN, we show that there still exists a strong monogamy relation:
\beq
\mathcal{D}_M^{A\rightarrow B}+\mathcal{D}_M^{A\rightarrow C} \leq \frac{1}{2}(1-a^2)=\mathcal{D}_M^{A\rightarrow BC}. \label{MoReMIN}
\eeq

The proof is as follows. First, note that any pure three-qubit state~(\ref{3qubit}) can be locally transformed into a standard form~\cite{AACJLT00,AAER01}
\beq
 \lambda_0\ket{000}+\lambda_1e^{i\phi}\ket{100}+\lambda_2\ket{101}+\lambda_3\ket{110}+\lambda_4\ket{111},\label{standard}
\eeq
with $\lambda_j\geq 0$ and $\sum_j\lambda_j^2=1$. Second, from the equalities $\tr{\rho_{AB}}^2=\tr{\rho_C}^2$and $\tr{\rho_{AC}}^2=\tr{\rho_B}^2$ for pure states, we obtain two useful relations:~\cite{CMHW16}
\begin{align*}
\tr{T_{AB}T^\top_{AB}}=1+2c^2-a^2-b^2, \\ 
\tr{T_{AC}T^\top_{AC}}=1+2b^2-a^2-c^2,
\end{align*}
where $a, b, c$ correspond to the module of local Bloch vectors $\mf{a}, \mf{b}, \mf{c}$ for Alice, Bob, and Charlie, respectively. Combining these results with Eqs.~(\ref{aneq0}) and~(\ref{identity}), we are able to calculate
\begin{align}
&\mathcal{D}_M^{A\rightarrow B}+\mathcal{D}_M^{A\rightarrow C}-\mathcal{D}_M^{A\rightarrow BC}\nn \\
=& \frac{1}{4}\left[2+b^2+c^2-2a^2-\frac{1}{a^2}\left(\mf{a}^\top T_{AB}T^\top_{AB}\mf{a} +\mf{a}^\top T_{AC}T^\top_{AC}\mf{a} \right)\right] -\frac{1}{2}\left(1-a^2\right)\nn \\
=& \frac{-2\lambda^2_0}{a^2}\left[4\lambda_1^2\lambda^2_2\lambda^2_3\sin^2\phi+ \left(\lambda_4(2\lambda^2_2+2\lambda^2_3+2\lambda^2_4-1)+2\lambda_1\lambda_2\lambda_3\cos\phi\right)^2\right ]		\nn\\
\leq & 0,
\end{align}
as claimed in Eq.~(\ref{MoReMIN}). 

As a by-product, Eqs.~(\ref{ordering}), (\ref{identity}) and~(\ref{MoReMIN}) imply  corresponding monogamy relations for negativity and geometric quantum discord:
\begin{align}
(\mathcal{N}^{A:B})^2+(\mathcal{N}^{A:C})^2&\leq (\mathcal{N}^{A:BC})^2=1-a^2, \label{MoReN} \\
\mathcal{D}_G^{A\rightarrow B}+\mathcal{D}_G^{A\rightarrow C} &\leq  \mathcal{D}_G^{A\rightarrow BC}=\frac{1-a^2}{2}, \label{MoReG}
\end{align}
while for quantum discord,
\beq
(\mathcal{D}^{A\rightarrow B})^2+(\mathcal{D}^{A\rightarrow C})^2\leq (\mathcal{D}^{A\rightarrow BC})^2=h(0).
  \eeq
Thus, using the monogamy relation~(\ref{MoReMIN}) for MIN, we provide an alternative way to derive the monogamy relations for squared negativity~\cite{OF07}, geometric quantum discord ~\cite{SAPB12}, and quantum discord~\cite{BZYW13}.

\section{Monogamy relations for multi-qubit states}\label{multiqubit}

One natural question arises whether the monogamy relation~(\ref{MoReMIN}) of MIN could be generalized to mixed states or even multi-qubit states. Unfortunately, the answer is negative. For example, consider the following mixed three-qubit state 
\beq
\rho_{ABC}=\frac{1}{2}\left(\ket{000}\bra{000}+\ket{111}\bra{111} \right).
\eeq
Then, it follows from Theorem 3 in~\cite{LF11} that we have
\beq
\mathcal{D}_M^{A\rightarrow B}=\mathcal{D}_M^{A\rightarrow C}=\mathcal{D}_M^{A\rightarrow BC}=\frac{1}{4},
\eeq
and thus $\mathcal{D}_M^{A\rightarrow B}+\mathcal{D}_M^{A\rightarrow C}>\mathcal{D}_M^{A\rightarrow BC}$. 

Even for $n$-qubit systems with $n\geq 4$, we can construct a counterexample
\beq
\rho_{ABCD\dots}=\frac{1}{2}\left(\ket{0000\dots}\bra{0000\dots}+\ket{1111\dots}\bra{1111\dots} \right),
\eeq
such that 
\beq
\mathcal{D}_M^{A\rightarrow B}=\mathcal{D}_M^{A\rightarrow C}=\mathcal{D}_M^{A\rightarrow D}=\dots=\mathcal{D}_M^{A\rightarrow BCD\dots}=\frac{1}{4}.
\eeq
This implies that no monogamy relation of MIN, similar to~(\ref{MoReMIN}), holds for multi-qubit states.

However, we can still derive a weaker monogamy relation of MIN for the general case in the sense that if the MIN between Alice and Bob is maximal, then either Alice or Bob must be in a product state with the rest parties, i.e., both $\mathcal{D}_M^{A\rightarrow C}$ and $\mathcal{D}_M^{B\rightarrow C}$ must vanish.

\subsection{Mixed three-qubit states}\label{mixed}

Generally, it is rather difficult to obtain an analytical form such as Eq.~(\ref{identity}) for the corresponding nonclassical measures for the partition $A:BC$. Instead, using the convexity of correlation measures acting on the states, we have a weaker monogamy relation of Bell-CHSH nonlocality~\cite{CH17}
\beq
\mathcal{M}^{A:B}+\mathcal{M}^{A:C}\leq 2, \label{tradeoff}
\eeq
for mixed three-qubit states. 

Then, following from the strength ordering~(\ref{ordering}), we can obtain a chain of monogamy relations of nonclassical correlation measures:
\begin{align}
&(\mathcal{N}^{A:B})^2+(\mathcal{N}^{A:C})^2,~~(\mathcal{D}^{A\rightarrow B})^2+(\mathcal{D}^{A\rightarrow C})^2 \nn \\
\leq & 2\mathcal{D}_G^{A\rightarrow B}+2\mathcal{D}_G^{A\rightarrow C}\nn \\
\leq & 2\mathcal{D}_M^{A\rightarrow B}+2\mathcal{D}_M^{A\rightarrow C}\nn \\
\leq & \frac{1}{2}\mathcal{M}^{A:B}+\frac{1}{2}\mathcal{M}^{A:C} \nn\\
\leq & 1. \label{chain}
\end{align}
This chain indicates that if the nonclassical correlation shared by Alice and Bob achieves the maximal value, then there is zero correlation between Alice (or Bob) and Charlie. Note that we may also use the convex-roof construction~\cite{CKW00} to derive tighter monogamy relations, but, the above trade-off relations require no optimization over all pure decompositions, which is much easier to verify.

\subsection{Multi-qubit states}

The monogamy relation~(\ref{tradeoff}) can be generalized as 
\beq
\mathcal{M}^{A:B}+\mathcal{M}^{A:C}+\mathcal{M}^{A:D}+\dots \leq 2\frac{\binom{n-1}{2}}{\binom{n-2}{1}}=n-1, \label{multi}
\eeq
for any $n$-qubit state $\rho_{ABCD\dots}$, either pure or mixed. Here we can pick up  $\binom{n-1}{2}$ groups, containing $A$ and arbitrary $X\neq Y$ from $B, C, D,\dots$, and thus sum $\binom{n-1}{2}$ inequalities of the form of Eq.~(\ref{tradeoff}) with respect to Alice, with each individual parameter appearing in total $\binom{n-2}{1}$ times. This relation is tight because it can be saturated by the generalized GHZ-states $\alpha\ket{00\dots0}_n+\beta\ket{11\dots1}_n$. 

We point out that the relation~(\ref{multi}) is different from the tradeoff relation for Bell nonlocality in~\cite{QFL15} where a tradeoff relation for all possible pairs is derived. Additionally, as we are constrained by quantum theory, the relation~(\ref{multi}) is stronger than the one derived within the no-signaling theory~\cite{T09,PB09}
\beq
\rt{\mathcal{M}^{A:B}}+\rt{\mathcal{M}^{A:C}}+\rt{\mathcal{M}^{A:D}}+\dots \leq n-1. 
\eeq

Consequently, the average of the maximal possible violation of a Bell-CHSH inequality, by Alice with one of  the other parties, is bounded by

\beq
\bar{\mathcal{M}} :=\frac{\mathcal{M}^{A:B}+\mathcal{M}^{A:C}+\mathcal{M}^{A:D}+\dots}{n-1}\leq 1. \label{average}
\eeq
As pointed before, this bound could be achieved  by the generalized GHZ-states. Correspondingly, the averages of other measures of nonclassical correlations satisfy
\begin{align}
\bar{\mathcal{D}}_M& :=\frac{\mathcal{D}_M^{A\rightarrow B}+\mathcal{D}_M^{A\rightarrow C}+\mathcal{D}_M^{A\rightarrow D}+\dots}{n-1}\leq \frac{1}{4}. \\
\bar{\mathcal{D}}_G& :=\frac{\mathcal{D}_G^{A\rightarrow B}+\mathcal{D}_G^{A\rightarrow C}+\mathcal{D}_G^{A\rightarrow D}+\dots}{n-1}\leq \frac{1}{4}, \\
\bar{\mathcal{D}} &
:=\frac{\mathcal{D}^{A\rightarrow B}+\mathcal{D}^{A\rightarrow C}+\mathcal{D}^{A\rightarrow D}+\dots}{n-1}\leq \frac{1}{\rt{2}}, \\
\bar{\mathcal{N}} &
:=\frac{\mathcal{N}^{A:B}+\mathcal{N}^{A:C}+\mathcal{N}^{A:D}+\dots}{n-1}\leq \frac{1}{\rt{2}}.
\end{align}
It is easy to find that the average amount of quantum correlations, shared by Alice and the other parties, is always upper bounded by half of the maximal amount for a bipartite system.

\section{Conclusions}\label{Discussion}

We have first studied the strength ordering of different measures of quantum correlations and found that for two-qubit states, the nonlocal effect MIN is always upper bounded by the maximal violation of the Bell-CHSH inequality given in Eq.~(\ref{MINH}). Then, we obtained a tight monogamy relation of MIN~(\ref{MoReMIN}) for all pure three-qubit states, which further gives an alternative derivation of known monogamy relations for squared negativity, quantum discord, and geometric quantum discord.  Finally, we could obtain a chain of weaker monogamy relations of various measures~(\ref{chain}) for general three-qubit states and the corresponding tradeoff relations~(\ref{multi}) for any multi-qubit state.

It is the monogamy of quantum correlations that at least in parts underlies their usefulness as resources in quantum information processing, especially in cryptographic protocols~\cite{E91}. Our results may provide insight into exploiting this usefulness. At the same time, it is still a rather complex problem to characterize and quantify nonclassical correlations in multi-party systems and thus it is also hoped that the further investigation of our work will help to expose the rich structures of quantum correlations. Finally, we may conjecture that the weak point that MIN based on the HS-norm in Eq.~(\ref{measurement}) is non-contractive under the CPTP maps on the local side, such as Bob's qubit, is helpful to obtain the strict monogamy relation~(\ref{MoReMIN}). Particularly, we find that the MIN based on the trace-norm $N_1$~\cite{MH15} without this weak point has no such monogamy relation for pure three-qubit states. For example, for the GHZ-state $\frac{1}{\rt{2}}\left(\ket{000}+\ket{111}\right)$, we have $N_1(\rho_{AB})=N_1(\rho_{AC})=N_1(\rho_{ABC})=1$, implying no monogamy relation exists for pure three-qubit states. It would be of interest to study whether MIN based on the relative entropy~\cite{XWL12}, the fidelity~\cite{MS17}, and two-side form~\cite{G13} satisfy similar monogamy relations or not.

\section{Acknowledgements}
	
	We thank Dr.~Michael Hall for helpful comments. S. C. is supported by the ARC Centre of Excellence
	CE110001027, and L. L is supported by National Natural Science Foundation~(NNSF) of China (Grant No. 61703254).

%
%






\bibliographystyle{apsrev4-1}

\bibliography{4qubit}

\end{document}